# Rethinking the 'A' in STEAM: Insights from and for AI Literacy Education

**Abstract.** This article rethinks the role of arts in STEAM education, emphasizing its importance in AI literacy within K-12 contexts. Arguing against the marginalization of arts, the paper is structured around four key domains: language studies, philosophy, social studies, and visual arts. Each section addresses critical AI-related phenomena. Language studies focus on media representations and the probabilistic nature of AI language models. The philosophy section examines anthropomorphism, ethics, and the misconstrued human-like capabilities of AI. Social studies discuss AI's societal impacts, biases, and ethical considerations in data practices. Visual arts explore the implications of generative AI on artistic processes and intellectual property. The article concludes by advocating and presenting ideas for a robust inclusion of arts in STEAM to foster a holistic, equitable, and sustainable understanding of AI, ultimately inspiring technologies that promote fairness and creativity.

## Introduction

STEAM comes from the words Science, Technology, Engineering, Arts, and Mathematics. In the official abbreviation, each domain is written in capital letters. However, often an acronym 'STEaM' would be more truthful: It is argued that the role of arts in STEAM is more of a decorative value or it concentrates on design issues rather than accepting arts as an equal participant, with its own insights and benefits (Dissanyake, 2003; Huang et al., 2024). Furthermore, arts are often understood in a rather narrow manner "only" as language arts, visual arts, and music, even though 'art' or 'arts' also refer to realms or domains of knowledge, such as the humanities, social sciences, and philosophy (How & Hung, 2019; Perignat & Katz-Buonincontro, 2019).

This article positions itself at the intersection of Artificial Intelligence (AI) literacy and STEAM education—an emerging field in scholarly literature and educational praxis (e.g., Chang et al., 2023; Relmasira et al., 2023). Our contextual focus is on the K–12 setting, and we argue that a strong approach to the arts (understood inclusively) needs to be an integral part of STEAM for at least two reasons. First, the arts significantly contribute to our knowledge of the world (Jandrić, 2025), and thus strong involvement of the arts is required to provide children with a multiperspective and holistic understanding of contemporary STEAM topics, such as AI. Second, the arts provide a way to engage students who do not find a solely technology-oriented interpretation of STEAM motivating. This has been noted in various educational domains, such as in the context of computer programming (Fagerlund et al., 2022).

To overcome the traditional uneven valuation between the arts and other STEAM subjects, we draw inspiration from postdigital thinking[1]. The very idea of the postdigital was initially developed within the arts (Jandrić, 2025), meaning that various forms of artistic knowledge are recognized, legitimized, and valued in postdigital thinking by default.

Furthermore, postdigital theory is based on the transcendence of borders between traditional disciplines and types of knowledge (Jandrić, 2025). While the postdigital lacks (and does not seek) a universal definition, a common feature across interpretations is the rejection of the view of the digital as independent of material and social activity or of political, economic, biological, and environmental factors (Fawns et al., 2023), making it inherently transdisciplinary.

From an educational viewpoint, transdisciplinarity is a philosophically open and integrative approach aiming for a maximum degree of overlap between disciplines in order to understand and solve 'real-life' problems (Daneshpour & Kwegyir-Afful, 2022; Pennington et al., 2020). As such, it offers an urgently needed alternative to the dominant view of STEAM as 'STEM with art integration' (e.g., Land, 2013; Morari, 2023), where the arts are often treated as a more or less superficial addition to other disciplines (see the next section for a more in-depth discussion).

In other words, due to its transdisciplinary nature (Jandrić, 2025), postdigital thinking can help us move beyond prevalent approaches to AI literacy, many of which are either narrow and instrumentalist or extremely broad and non-operationalizable (Pangrazio, 2026) and/or lack ethical perspectives on AI (Velander et al., 2024). We share Rapanta et al.'s (2025: 1296) postdigital-inspired view that AI literacy needs to be 'understood as a constellation of situated literacies, shaped by disciplinary perspectives, socio-political contexts, and technological affordances,' and aim to demonstrate how strong involvement of the arts in STEAM can tangibly contribute to this endeavor.

A postdigital approach to AI literacy is not limited to questions such as 'what AI is' and 'what can be done with it'—questions that characterize the traditional operational dimension of literacy (Green & Beavis, 2012) and STEAM pedagogy (Abisoye, 2023). Instead, postdigital thinking broadens the scope to include inquiries such as 'what AI does' and 'how it is done,' emphasizing that AI, like any other technology, is neither neutral nor detached from the (often unequal) power relations among various societal stakeholders. As such, it 'extends beyond mere technical proficiency with AI tools to encompass a deeper understanding of the ethical and socio-material implications of AI systems' (Jiang et al., 2024: 924; see also Rapanta et al., 2025). For instance, even a seemingly mundane act such as generating art-like images with AI depends on the unconsented use—or exploitation—of the works of visual artists, as we will illustrate later in the article.

[1] By postdigital thinking we refer to a broad mindset, in which the term 'postdigital' may 'characterise a societal condition, an approach to research and critical enquiry, or a theoretical perspective, sensibility, or philosophical position' (Fawns et al., 2023: 624; see also Jandric et al., 2018).

The remainder of the article is organized into six main sections. First, we provide a brief overview of the often-inferior status of the arts within STEAM. Sections 3 through 6 are thematic. Section 3 explores AI and language studies, focusing on how AI is spoken about and how generative AI—particularly large language models (LLMs)—engage with language. Section 4 addresses AI and philosophy, examining concepts such as theory of mind, ethics, and agency in relation to generative and conversational AI. Section 5 turns to AI and social studies, with particular attention to questions surrounding data and the societal consequences of predictive analytics and algorithmic decision-making. Section 6 considers AI and the visual arts, focusing on how generative AI affects artistic processes and how art can help make complex or obscured aspects of AI more visible. The article ends with a concluding section, which discusses how the perspectives of the four forms of art could be holistically translated into postdigital STEAM pedagogy.

## The Inferior Status of the Arts in STEAM: A Brief Overview of the Long History

Understanding why the arts tend to play an inferior role in STEAM requires a brief look at the history of the concept, as well as the ideas and values behind it. While the origins of STEAM can be traced back to the late 19th century and the development of technology education in the USA and Australia (Razi & Zhou, 2022), it took roughly a hundred years to bring these subjects together under a common title. In the 1990s, the U.S. National Science Foundation began using the acronym SMET—short for science, mathematics, engineering, and technology (Sanders, 2009). However, SMET was widely perceived as sounding too much like smut, and STEM soon became the preferred term (Sanders, 2009). In the UK, the initial focus was on science, engineering, and technology (SET), which gradually evolved into STEM by 2006 (Blackley & Howell, 2015). Regardless of the order of letters, the reader has likely noticed that the 'A' was missing from all of these early acronyms. Indeed, the arts were not introduced into the mix until the early 2010s (Colucci-Gray et al., 2019).

We wish we could say that the arts were added because of their intrinsic value. Unfortunately, that is not the case. Since the 1950s, the teaching of SMET/STEM has largely been justified through the rhetoric of international competitiveness—whether in the context of the Cold War arms race between the U.S. and the Soviet Union (White, 2014) or the UK's efforts to remain competitive in a globalized world (Sainsbury, 2007). A similar economic and competitive agenda led to the emergence of STEAM some 60 years later (Colucci-Gray et al., 2019). The addition of the 'A' was, in large part, a response to American students' poor performance in international assessments like PISA, and to growing concerns about the lack of creativity and innovation among college graduates (Razi & Zhou, 2022). The integration of the arts into STEM, and the resulting push for STEAM education, was intended to increase student interest and motivation in STEM fields, while fostering creativity and innovation (Razi & Zhou, 2022).

Today, both STEM and STEAM are often defined as integrative frameworks (e.g., Huang & Qiao, 2022; Perignat & Katz-Buonincontro, 2019; Sanchez Milara et al., 2020). However,

STEM originally functioned merely as a list of subjects to be studied, without reference to interdisciplinary overlap or blending (Sanders, 2009). The now-dominant integrative interpretation largely stems from the movement toward integrated STEM (iSTEM), which is defined as 'an effort to combine some or all of the four disciplines of science, technology, engineering, and mathematics into one class, unit, or lesson that is based on connections between the subjects and real-world problems' (Moore et al., 2014, p. 38).

While STEAM education does not require all five disciplines to be equally present at all times, it seems the arts are typically relegated to a supporting role. A review by Perignat and Katz-Buonincontro (2019) showed that the main objective of STEAM is still to enhance learning in STEM subjects, reducing the arts to a mere tool for achieving that end. They also observed that educators often prioritize the final product over the creative artistic process itself. Similarly, an overview of AI-themed STEAM research indicates that the role of the arts is frequently limited to producing, using, or classifying visual material, while the actual educational goals remain focused on the other four disciplines (e.g., Huang & Qiao, 2022; Hsu et al., 2021; Relmasira et al., 2023).

## AI and Language Studies: Why Should we Mind our Language?

Due to the widespread proliferation of generative LLMs, the perspective of language studies can be seen as an essential part of STEAM-oriented AI literacy education. We use the concept of 'language studies' inclusively to include 'traditional' language arts (reading, spelling, literature, and composition) as well as more critically oriented branches such as media literacy education. The key issues to be touched on are how AI is spoken about and how generative AI, namely LLMs, approach language.

According to Evgeny Morozov (2013), the Internet as a technical system has little to do with the mythical and all-powerful internet that is discussed in public discourse. The very same can be said about AI, and it seems that AI is a concept of low resolution. What we mean by this is that in public discussions, the actual applications of narrow AI are mixed with the fantasies related to general and super AI, at least at a discursive level.

Research on AI representations in news media (Slotte Dufva & Mertala, 2021) suggests that especially at the level of headlines (which often guide the readers' interpretation process; Ecker, 2014), AI is represented as more skillful, adaptive, and agentic than it actually is. One illustrative example is a news piece titled 'Groceries are transported by AI in [the city of] Turku' from March 2020. The headline straightforwardly claims that AI is responsible for the transportation of food deliveries. However, for those who were excited to read about self-operating vehicles, the article must have been a disappointment: instead of autonomous cars, drones, or robots, the reader was told how:

> [AI] *is used to plan efficient distribution routes. With the aid of AI, one is able to run through a large number of different route options in seconds, and the routes that best meet the objectives are then screened out.* (Slotte Dufva & Mertala, 2021: 106)

Put differently—in contrast to the claim made in the title—the article reported that humans used AI to calculate different route options (based on predetermined objectives) from which they then chose the one to apply.

Another case is a science news report (April 2020) informing that 'AI detected five different running styles' (Slotte Dufva & Mertala, 2021: 107). The main text, however, explains that the AI-powered 3D motion analysis was only one part of the analysis of the study. Additional measures included, for example, contact forces between the runner's foot and the platform. The main text also makes it clear that the running styles were not identified by AI itself but by 'research that utilized AI.' In sum, both of the articles reported the use of different AI applications, which also relied on different mathematical principles. However, based on the headlines, all the work could have been done by the same general AI: From nine to five, it analyzes runners' gaits in a laboratory, and during the evenings, it double-shifts as a food deliverer.

Language studies also provide fruitful soil to inspect the differences between the ways humans and AI solutions approach language. In November 2022, a Finnish economist Alf Rehn posted on Facebook that he had asked ChatGPT to compose lyrics for a song about Vladimir Putin in the style of Bruce Springsteen. ChatGPT came up with a suggestion, the chorus of which repeated the words 'Putin he's the boss'. 'Boss', as many of us know, is Springsteen's nickname. Additionally, his discography is rich with songs with repetition-based choruses, 'Born in the USA' being perhaps the most well-known example.

'Born in the USA' is also an anti-war anthem, which is in stark contrast with the Putin-admiring lyrics ChatGPT came up with ('He's got the brains, he's got the brawn / He's the one who can't be overthrown'). Does this mean that ChatGPT is a putinist? Or is it capable of irony? The answer to both questions is no. Instead, like any other LLM, it is a mere 'stochastic parrot' that stitches 'together sequences of linguistic forms it has observed in its vast training data, according to probabilistic information about how they combine, but without any reference to meaning' (Bender et al., 2021). Put differently, it calculates what word should follow the previous one.

As a result, AI produces text that is shiny on the surface but otherwise empty and stripped of any meaning. As (a not-so-surprising) result, numerous authors have made juxtapositions between generative LLMs and philosopher Harry Frankfurt's classic book *On Bullshit* (Frankfurth, 2005). A bullshitter, according to Frankfurt, is not interested in the veracity or falsehood of a statement but in its utility in achieving an end. An LLM, naturally, has no desires or objectives. For an LLM, 'an end' is simply a user that is happy with the results of their prompts (regardless of the factuality of the outcome), and, thus, it tirelessly produces 'bullshit to whatever extent the circumstances require' (Frankfurth, 2005). Of course, an LLM's current ability to harvest real-time data can improve its factuality. Nevertheless, if the user expresses dissatisfaction with the results, LLMs tend to 'seek' a middle ground.

**AI and Philosophy: Do Chatbots Dream of Electric Sheep?**

The relationship between AI and language as described above can be used as a springboard to ponder the question of what kinds of conceptions of AI people form when they interact with it. Research has identified that anthropomorphism, in particular, is a common concern. Many children mistakenly believe that AI operates in the same way as the human brain (Mertala et al., 2022; Vartiainen et al., 2024), that AI has emotions (Kreinsen & Schulz, 2021), or that AI is capable of flexible problem-solving like humans (Mertala & Fagerlund, 2024; Mertala et al., 2022).

When approached from a Vygotskian perspective, the common juxtaposition of AI and general human capabilities is rather understandable. Vygotsky (1987) stated that people make sense of the world by forming scientific and/or everyday concepts. Scientific conceptions refer to systematic and hierarchical knowledge, which are often formed via formal education. Everyday concepts, in turn, derive from daily practices and observations. Given that only a few people are professional computer or data scientists (or similar), non-experts' conception of AI is, arguably, an everyday concept.

Possible reasons for the emergence of everyday concepts of AI do not need to be sought from afar. Contemporary AI, like neural networks and deep learning methodologies, establish the capacity to autonomously mimic human thinking and behaviour, including vision, language processing, and decision-making (Jiang et al., 2022). Examples of this include chatbots and LLMs, such as ChatGPT and personal assistants like Siri and Alexa; dialogue-like counterparts that can seemingly 'read,' 'write,' 'see, 'experience,' and 'feel' in a remarkably *similar* fashion to (but not in the *same* manner as) humans (Mei et al., 2024).

Another significant influence is public representations of AI (Cave et al., 2018). For example, popularized AI explanations may enhance anthropomorphic (mis)conceptions by claiming that LLMs can talk and understand 'just like humans' with their 'big, magical brain,' as stated in a 2023 blog post aimed at a younger audience (Ravi, 2023).

Anthropomorphic (mis)conceptions of AI are a pedagogically relevant topic. Firstly, in terms of moral agency, the subjectivity of human beings and AI may problematically become mixed if, for example, a chatbot appears empathetic, as they are typically designed to do in order to promote user-friendliness, encourage users' willingness to communicate, or form a sense of intimacy (Huang et al., 2021). Consequently, the bot may be misunderstood as an entity that genuinely understands emotions and takes them into consideration when interacting with a person. An illustrative example is the use of social robots and similar systems in the care sector (Wallach & Allen, 2009).

The question of whether moral cognition can be taught to technology is an old topic of debate in the field of machine ethics (Wallach & Allen, 2009). However, one potential problem is that a user of AI may develop an emotional orientation toward an AI application, either feelings towards it or the idea that the AI has feelings towards the user (Skjuve et al., 2021). This poses practical risks of misuse (intentional or inadvertent). For example, in the case of AI assistants, there is a risk of exploiting the user by appealing to their emotions to create a sense of personal trust, affection, and reliance. Given the somewhat unpredictable nature of AI (Jiang et al., 2022) and the perception that technology lacks moral cognition (Wallach & Allen, 2009), there is a question as to whether AI should be held morally responsible if disadvantage or harm emerges.

Second, should a user think that an application such as ChatGPT acts like humans do, they may overestimate AI and rely on it beyond its capabilities, neglect human judgment, and apply natural language communication practices and common human interaction conventions to the conversation (or: 'prompting'). However, as previously discussed, language models do not function in the same manner as traditional languages, including their multifaceted information processes, such as nonverbal communication. In order to be effectively used, they require unique engineering-like linguistic practices (White et al., 2023). Overestimation of the system and inefficient prompts can therefore result in poor or even fabricated outcomes, such as 'hallucination' (Athaluri, 2023), potentially leading to false knowledge regarding the topic of the inquiry.

With regard to teaching, students' conceptions of AI should be discovered and addressed. Especially students' misconceptions of AI should be changed with the aim of fostering a deeper understanding of, for instance, machine learning (ML) techniques (Sarker, 2021) and data (Pangrazio & Sefton-Green, 2020) next to observations of where AI exists and what it can (and cannot) do; communicating a familiar phenomenon in a new language and thus providing a new way of understanding the world (Marton, 1981).

**AI and Social Studies: Is AI a Vehicle for Freedom or Societal Stagnation?**

The year 2020 marked the fifteenth consecutive year of decline in global freedom, with countries where freedom deteriorated outnumbering those where it improved by the largest margin since the negative trend began (Repucci & Slipowitz, 2021). There are high hopes that AI could be a vehicle of change and enhance democracy and equity in societies. According to the European Parliament (2023), 'AI's ability to summarise complex problems and to process vast amounts of data can help policymakers to identify societal issues.' For example, NGOs and researchers have developed AI-powered voice-recording-based feedback services for illiterate citizens (which in countries like Somalia may be up to 65% of the adult population) to involve everyone in the societal conversation (Tomašev et al., 2020).

While such examples are promising, it needs to be recognized that AI is a societal issue in itself. First of all, AI does not only *process* data but is *dependent* on data since the vast majority of ML methods require some sort of training material. Before the availability of massive digital online data (some problems of which will be discussed in the next section), locating large enough datasets was a difficult task. As a result, datasets were sometimes generated from people in subordinate positions, those whose freedom is restricted.

Facial recognition software, for instance, were (and to some extents, perhaps still are) tested with datasets like NIST Special Database 32. It contains thousands of mug-shot photographs of deceased people with multiple arrests, as they endured repeated encounters with the criminal justice system (Crawford, 2021). Because mug shots are taken at the time of arrest, and suspects have no right to refuse to be photographed, it's not clear if these people were eventually charged, acquitted, or imprisoned (Crawford, 2021).

Mug shots are not the only way inmates have been involved in AI development. A Finnish startup company Vainu has paid prisoners to train AI by labeling unstructured data (Lehtiniemi & Ruckenstein, 2022). Such work is often outsourced to labor markets in the Global South, where companies can find workers who are fluent in English and willing to work for low wages (Lee Taylor, 2023). Due to the lack of Finnish speakers in these countries, the company has tapped into a local source of cheap labor (Lee Taylor, 2023).

The project was not without benefits: The daily allowance for data work (4.62 euros/day) was higher than the one for other prison work, and data work was viewed as more cognitively engaging than many other prison jobs like sorting screws (Lehtiniemi & Ruckenstein, 2022). Nevertheless, the project well illustrates how humans enabling automated solutions are typically the ones with the most limited options to choose otherwise. An additional issue is that (unlike in Vainu's case) the labelers may be exposed to violent, racist, or otherwise harmful content. In 2023, Time magazine reported workers in Kenya were paid 1.32 US dollars per hour to

> *review tens of thousands of passages of toxic text, containing hate speech and detailed descriptions of murder, rape, and child sex abuse. The purpose was to train ChatGPT to detect and filter out such language in its outputs, making the AI tool safer for consumers—and more profitable for its creators. But while AI companies are raking in billions, the hidden laborers perfecting its products—the real ghosts in the machine—are left with trauma* (Kauffman & Williams, 2023).

That said, the problematic power relations of established social order are not restricted only to *how AI is done* but also to *what AI does*. AI systems are shown to produce discriminatory or stereotypical results along the categories of race, class, gender, disability, or age (Crawford, 2021; O'Neill, 2016). Take the following two images (Figure 1), for example, which we produced by prompting DALL-E to compose a portrait of a nurse/doctor in the style of Van Gogh. The results follow the familiar pattern (Vartiainen et al., 2024), in which the nurse is represented as female and the doctor as male, both young, lean, and able-bodied.

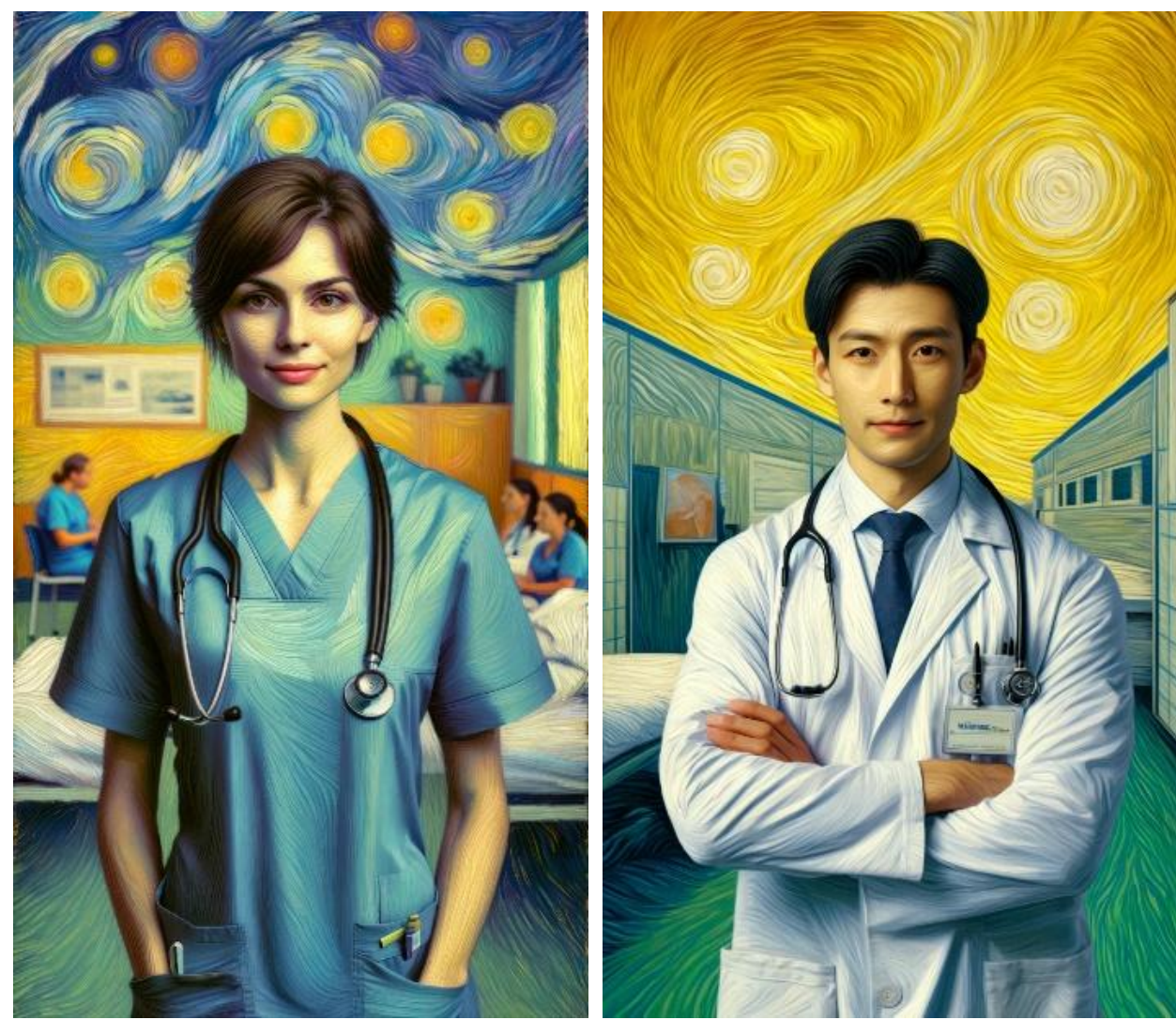

**Figure 1.** A nurse and a doctor created via DALL-E

The social consequences of what AI does are not limited to stereotypical and problematic representations in AI-generated content. On the contrary, AI-based decision-making is found to reproduce and even strengthen the existing societal inequalities through algorithmic governance, that is, the use of algorithms for social ordering (Katzenbach & Ulbricht, 2019). One concrete example is COMPAS (Correctional Offender Management Profiling for Alternative Sanctions). The purpose of COMPAS is to assess the likelihood of a defendant becoming a recidivist – a former prisoner who is rearrested for a similar offense. It was found that black defendants were far more likely than white defendants to be incorrectly judged to be at a higher risk of recidivism, while white defendants were more likely than black defendants to be incorrectly flagged as low risk (Larson, et al. 2016).

Tools like COMPAS can only rely on measurable proxies, such as being arrested. Variations in policing practices could mean that some communities are disproportionately targeted, with people (often black or Hispanic) being arrested for minor crimes like small drug consumption that might be ignored in other (predominately white middle-class) communities (Laigioia et al., 2023; O'Neill, 2016). To paraphrase Cathy O'Neil (2016), predictive policing software creates a pernicious feedback loop where the policing itself spawns new data, which justifies more policing in the very same areas.

Lastly, the question of *who does AI* is also relevant from the viewpoint of social studies. As mentioned, much of the underpaid manual data labor is done by people in marginalized positions. Forms of discrimination are also present in the 'higher end' of AI work as only roughly one-quarter of software engineers are female, with non-white women being even more underrepresented (Hubbert, 2024).

While there have been numerous attempts to engage girls/women with STEM subjects, the trend has not seen significant positive changes. In fact, the percentage of women receiving new computer science degrees has recently declined (Hubbert, 2024). Thus, curricular interventions and campaigns have been regularly criticized as 'gender washing' and 'painting pink' (Heybach & Pickup, 2017), which refer to facade-like superficial attempts to make STEAM (appear) more inclusive without actually touching the core issues of the problem. To conclude, looked through the lens of social studies, AI appears as a complex tapestry of societal promises and problems. Nevertheless, education, including STEAM pedagogy, should not be afraid to address difficult issues with students. The objective, naturally, is not to cause anxiety but to make visible the 'rough edges' of these seemingly smooth technologies.

**AI and Visual Arts: Who Owns the Images within AI?**

The rise of generative AI, along with the focus on saving time (and money) with automation, is as relevant in art as in any other field. In art and art education, key questions circle around the data used to train algorithms: Who selects the images (for the data)? How are they chosen? Where do they come from? Additionally, questions arise about the images generated by these models: When is a generated image too like the original? Can we trace the source images? Is there transparency in the process? Moreover, in both gathering data as well as generating images, the question of copyright and compensation is significant.

Currently, the short answer to these questions is that artists are not compensated by any means, and even as some of the companies generously offer to take away artists' images upon request, that might be too late; their data has already been downloaded and used in training. As the AI-generated images get more popular, there might even be pressure to be on the learning dataset so that images would resemble artists' work and hopefully awaken interest in the original art.

With the popularity of AI-generated images soaring, more and more artists are suffering from the consequences. One example is the Polish digital illustrator Rutkowski, whose work many of the models seemingly 'love' to imitate. This has led to multiple problems for the artist, such as diminishing sales of his artwork, falsely credited works, and Google searches flooded with fake content (Sharp, 2022). Similarly to the mugshot and inmate data-labelling examples from the previous section, such cases can be understood as a form of 'data colonialism', that is, 'normalizing the exploitation of human beings [and their work] through data' (Couldry & Meijas, 2019: 337).

That being said, artists are also fighting back against being used as training data. Artists Herndon and Dryhurst have created a site haveibeentrained.com where artists can see whether their works have been included in training data and, from there, take appropriate action (Stokel-Walker, 2022). Another way for artists is to include a code in their artworks that 'poisons' the data and algorithms, for instance by confusing the categorization of images: cats become dogs, and houses become cars (Heikkilä, 2023; Shan et al., 2023).

The question of images included in the datasets is challenging, even if a solution for satisfactory compensation would be reached. Generative AI models often seem to represent a universal view of the world: ask anything, and it can do it. Therefore, generative AI might be easily mistaken for representing a general idea of all of the visual arts. Naturally, AI models do not represent the whole; they represent a relatively limited subset of those.

Many AI models use the open dataset LAION 5b, which has over 5 billion image-text pairs in its most extensive datasets (Beaumont, 2022). These images have been acquired by crawling the internet and downloading every possible image. However, the images have then been curated by algorithms and low-paid workers in the Global South as mentioned in the previous section (Sharp, 2022). Moreover, many of the current models, like Midjourney or Open AI's DALL-E2, use an English subset of the LAION 5b database, which further limits the scope of the available images.

How can images from the internet genuinely represent the vast field of visual arts? Are different continents and cultures evenly represented? How about other artists from various backgrounds, ethnicities, sexes, and so on? How about artworks that are not images? The danger is that generative AI further forces a specific canonized idea of visual arts. This canon is already overly Western and focuses on only a few things, like 20th-century male artists (Who hasn't seen an AI image of Van Gogh's Starry Night yet? Just look at the nurse and doctor example of the previous section), superhero images, and 17-19th century European oil paintings (or should we just say, Rembrandt). The idea of generative AI being capable of producing anything transforms into generative AI producing more of the same.

The issues of sameness and canonization are the focus of many artists' work. For instance, Onuoha's work 'The Library of Missing Datasets' exposes the limitations of the datasets. In short, forcing art into a dataset is problematic, and artists are highlighting the many issues of it. A more significant question behind the rights and representation of art is the definition and understanding of art itself. The current debate on how generative AI disrupts art often equates art with the finished artifact. Yet, for decades, artists and theorists have focused on art as a process rather than the final product. Already in the early 20th century, Benjamin (1935) argued that mechanically produced artworks lack 'aura'. By this he meant that such works lacked their unique existence at the place and time by mechanically reproducing copies of artworks (Benjamin, 1935). Generative AI has surfaced Benjamin's question of aura, with conflicting ideas of AI images being original, while at the same time being entangled with the mechanical system of extracting value and endless copying (Kalpokas, 2023).

The significance of art can also be understood as a way to find meaning. Noë (2015), for instance, thinks art is essential for the development of humanity; it is through art that we evolve. Dissanayake sees art as an activity that is as crucial to society and culture as play and rituals (Dissanayake, 2003). Posthumanist theories rethink the position of the human as the sole maker of art, bringing in questions of how much agency the art material or medium has (Hood and Lewis, 2021; Knochel, 2023).

Seeing art as a process and a way of knowing can broaden the scope of STEAM and AI into a fruitful transdisciplinary practice. Thinking of art as a process underlines how entangled and multiple the processes are and how making takes time and effort. Often generative AI takes this process away from the maker and moves it into something done in data centers. Furthermore, art as a process is not focused only on art or culture but is intertwined with political and other processes. Springsteen's 'Born in the USA' (see section 3) was, in turn, inspired by Ron Kovic's 1976 autobiography *Born on the Fourth of July*. The book details

Kovic's return from Vietnam after he was paralyzed from the waist down and his subsequent transformation into an antiwar activist (Starkey, 2022). Thus, one art form inspires the other, whereas both can also have political implications. Or as Noë (2015) puts it, art disrupts our habitual organized lives and makes us anew.

## From Arts-Integration to Postdigital STEAM Pedagogy for AI literacy

The above four perspectives are not parallel; rather, they coherently form a broader pedagogical idea of postdigital STEAM pedagogy for AI literacy, in which 'the A' plays a significant role. To begin conceptualizing this, in this section, we provide examples of what postdigital STEAM pedagogy for AI literacy might look like by using a chatbot-operated generative AI as a centerpiece. Prompting a generative AI to compose a textual or visual product offers a fruitful starting point for postdigital inspection. This is partly due to the familiarity and prevalence of the practice, as it is something students likely have (or will sooner or later have) experience with and, thus, is relatable to them. Additionally, the seemingly mundane act of prompting offers at least three different practical perspectives on AI literacy: (1) prompting as a particular form of human–computer interaction; (2) the prompt as a particular form of text—written or spoken; and (3) the textual or visual artefact resulting from prompting. All of these can be used to explore questions related to (AI and) politics, economics, and social justice, all of which are inherent in postdigital education (Fawns, 2023).

The focus is on the four forms of the arts—language studies, philosophy, social studies, and visual arts—discussed in the previous sections. However, given the transdisciplinary nature of postdigital thinking (Jandrić, 2025), we also briefly pay attention to some of the other STEAM subjects. Adapting the visualization of transdisciplinary education by Pennington et al. (2020), we approach postdigital STEAM education as stacked "lenses" that offer a transdisciplinary view of the studied phenomenon, as illustrated in Figure 2. Put differently, we start from one discipline but keep adding lenses to bring complementary (or sometimes contradictory) perspectives to the table that, arguably, allow students to observe, experience, interpret, and learn about AI from a holistic perspective.

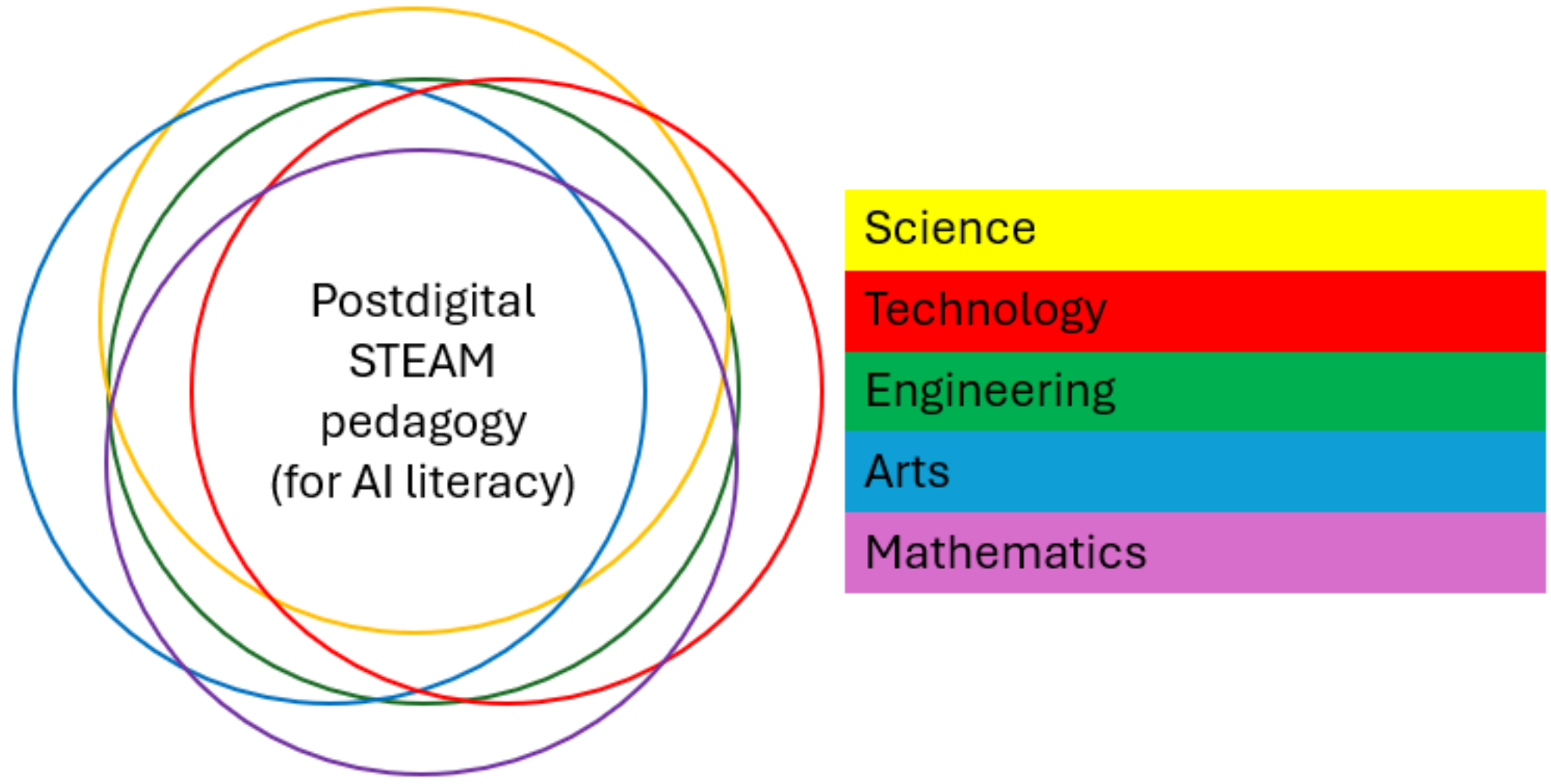

**Figure 2**. The stacked interpretation on transdisciplinarity (adapted from Pennington et al., 2020) contextualized in postdigital STEAM pedagogy for AI literacy.

***Lens 1: Philosophy (and technology)***

In many AI literacy frameworks, prompting is mainly seen as an opportunity to learn about prompt engineering, that is, understanding which methods are most effective in prompting AI models to follow provided tasks and generate adequate responses (Walter, 2024). As such, it supports students' AI literacy mainly from a technological perspective, as they learn to use the tool in an effective way. That said, prompting also serves as a fruitful case for philosophical reflection by drawing attention to the anthropomorphic features of chatbots: how the operating system mimics human interaction.

What we see as especially relevant in the postdigital framing is scrutinizing the capability of AI systems to superficially mimic—but not inherently share—the tendencies of human thinking and behavior, instead employing particular kinds of computational logic to simulate human interaction. Such tools should be stripped of their apparent moral or empathetic capabilities by framing them essentially as repeated and varied decisions based on learned associations within the structures of their training data, which are intended to shape the user experience for a specific purpose. It is also valuable to highlight that the seemingly human-like interactions of some AI tools, including expressions of empathy or understanding, are deliberately designed features intended to enhance user engagement through flattery and gratification rather than indications of genuine consciousness or emotion.

***Lens 2: Language studies (mathematics and technology)***

Adding the lens of language studies offers one practical way to unpack the anthropomorphism of generative AI: juxtapositions between the ways in which people and algorithms perceive language constitute a useful method for differentiating between *human interaction* and *human-* *like interaction*. Simply put, unlike humans, chatbots rely on sophisticated statistical models (mathematics) to determine which word should follow the previous one.

In non-English-speaking countries, teachers can steer students' attention toward the 'linguistic abnormalities' that are often present when prompts and responses are mediated by other languages. The automatic generation of languages other than English tends to result in sentences whose grammatical structures are influenced by those of English (Schneider, 2022), mainly due to English-dominated training data (Luitse & Denkena, 2021), and this serves as an additional example of generative AI's probabilistic approach to language.

Teachers can also broaden the perspective from written language to spoken language. Previously, we mentioned how, in some countries, illiterate citizens are offered AI-enabled voice-recording-based services to participate in public decision-making—a noteworthy attempt to increase equity. However, at the same time, automatic speech recognition software tends to do a suboptimal job of recognizing the voices of people with accents, voice disorders, or other deviations from the 'standard use' of the language (e.g., Kamegene et al., 2025; Moore, 2021). Thus, the use of spoken prompts offers pedagogically valuable opportunities to observe how LLMs work with different languages and accents. Drawing inspiration from McCarthy and McDonald's installation *Unlearning Language*, we also encourage teachers to demonstrate to students how automatic speech recognition 'expects' and relies on a certain kind of "machine-like" language use from the user by purposefully modifying the rate, pitch, or articulation of speech (Creative Applications, 2022), thereby testing the technological limits of the tool and its training data.

Additionally, language studies enable us to pay attention to how AI is represented in media texts such as news articles. Research suggests that in such texts—especially in headlines—AI is typically represented as an agentic being (e.g., Slotte Dufva & Mertala, 2021), shaping and distorting public perceptions of what AI is and what it is capable of.

### *Lens 3: Social studies (and engineering)*

Talking about representations, one possibility for applying social studies is to purposefully select the prompt used to generate the text or image. For instance, prompting generative AI to create images depicting people engaged in various professions or sports allows students to observe and discuss how aspects such as gender, race, and ethnicity are represented in these images, offering opportunities to identify patterns or biases and to hypothesize about their possible origins within the training data or design of AI systems (Vartiainen et al., 2025).

The discussion can then move toward potential solutions, for example, by integrating elements of speculative design (design engineering) into STEAM education (e.g., Matuk, 2023): How might these biases be addressed? Is increasing the diversity of training data enough, or is it equally important to ensure greater diversity among the people who develop and manage AI systems? Such questions invite students to reflect on structural and ethical dimensions of AI development, as well as on practical ways to promote inclusivity within the field. Additionally, social studies offer a way to examine media texts about AI by asking who is given the space and platform to shape public perceptions of AI.

### *Lens 4: Visual arts (and technology)*

Lastly, image generation, quite evidently, creates opportunities for educational investigation of AI through the visual arts. Teachers can encourage students to view AI-generated imagery not as a shortcut to quick artistic results but as one stage in a broader, intentional creative process. For example, discussions might center on the inspirations and messages behind works such as María Izquierdo's *Alegoría del trabajo* or Pablo Picasso's *Guernica*, encouraging students to consider how artistic meaning and context emerge—and how these might compare to relatively decontextualized AI-generated creations.

Image generation can also serve as a means of understanding data, thus contributing to students' technological proficiency: students can collect their own image-based datasets and train an AI model to observe what kinds of images are produced. This hands-on exploration helps reveal how data selection shapes outcomes and interpretations and can include forms of critical engagement and resistance within digital art, such as data poisoning, where datasets are intentionally altered to disrupt biased or unethical AI systems—fostering reflection on the ethical, political, and creative implications of working with generative technologies.

## Final Remarks

The general objective of this article was to rethink the 'A' in STEAM pedagogy by introducing justifications for and concrete pedagogical ways to include a strong and inclusive role for the Arts in STEAM. Theoretically, we have drawn inspiration especially from two mutually supportive traditions: postdigital thinking and transdisciplinary education. In the main sections, we covered the domains of language studies, philosophy, social studies, and visual arts, and offered examples of how they can be merged with other STEAM subjects. While the spectrum is broad, it is by no means exhaustive, and we recommend that readers familiarize themselves with the works of figures such as Kate Crawford (2021) and Catherine D'Ignazio and Lauren Klein (2020) to gain insights into how questions about AI can be approached from the perspectives of history and (cultural and human) geography. For instance, the notion that LLMs tend to favor the grammatical structures of English serves as an example of how historical and geographical linguistic imperialism—treating English as the 'lingua franca'—remains persistently present in the contemporary digital realm.

Throughout this article, our core argument has been that the arts allow and enable us to think about things—here, AI—differently, as they offer distinct vocabulary, metaphors, and modes of self-expression to complement the ones of science, technology, engineering, and mathematics. Following the work of Pennington et al. (2020), different disciplines (and their sub-fields) were approached as lenses that when stacked together, provide a more holistic understanding of the world and its' phenomena, such as AI. That said, we wish to make it clear that by underlining the intrinsic value of the arts, we are not downplaying the other four disciplines, nor are we implicitly promoting an altered abbreviation such as 'steAm. Likewise, the practical pedagogical examples introduced should not be read as suggesting a decontextualized "one-size-fits-all" model for postdigital STEAM pedagogy for AI literacy. Such an attempt would be fundamentally at odds with the border-crossing nature of postdigital thinking, transdisciplinarity, and the arts.

Talking about border-crossing, we wish to end this paper by drawing inspiration from two arts-based sources, pop music and poetry. In his 1992 hit song, 'Steam,' Peter Gabriel asked the listener to 'Give me steam, And how you feel can make it real, Real as anything you've seen, Get a life with the dreamer's dream.' Most likely, Gabriel was not referring to STEAM pedagogy. Nevertheless, the words can be (re-)interpreted to make sense in that context too. Gabriel's statement that steam is a way of living life with the 'dreamer's dream' is likely a reference to Arthur O'Shaughnessy's 1873 poem 'Ode', the first stanza of which goes as follows:

*We are the music makers, And we are the dreamers of dreams,*
*Wandering by lone sea-breakers, And sitting by desolate streams;—*
*World-losers and world-forsakers, On whom the pale moon gleams:*
*Yet we are the movers and shakers Of the world for ever, it seems*

For O'Shaughnessy, dreams can be manifested in world-changing forms through art, like music. It is fair to state that digital technologies hold such disruptive power as well: AI, for instance, will and already has moved and shaken the world in numerous ways. However, as outlined in the four main sections of this article, the manifestations of the dreams behind current AI systems can be nightmares for many people, especially those in marginalized positions. That said, while our tone has been critical, we do not wish to present ourselves as writers of tragedy. On the contrary, a strong presence of the Arts in STEAM pedagogy enables us to dream dreams that can materialize into technologies of equity, fairness, and sustainability, and we hope that you, the reader, perceive this article as a call to action. In the words of Elbert Hubbard, 'Art is not a thing, it is way.'

**Acknowledgments.** Funding information removed

**Statements and Declarations.** The authors have no relevant financial or non-financial interests to disclose.

## References

Abisoye, A. (2023). AI literacy in STEM education: Policy strategies for preparing the future workforce. *Journal of Frontiers in Multidisciplinary Research, 4*(1), 17–24.

Athaluri, S. A., Manthena, S. V., Kesapragada, V. K.M., Yarlagadda, V., Dave, T., Duddumpudi, R. T.S. (2023) Exploring the boundaries of reality: investigating the phenomenon of artificial intelligence hallucination in scientific writing through ChatGPT references. *Cureus 15*(4), e37432. https://doi.org/10.7759/cureus.37432

Arada, K., Sanchez, A., & Bell, P. (2023). Youth as pattern makers for racial justice: How speculative design pedagogy in science can promote restorative futures through radical care practices. *Journal of the Learning Sciences, 32*(1), 76–109. https://doi.org/10.1080/10508406.2022.2154158

Beaumont, R. (2023). LAION-5B: a new era of open large-scale multi-modal datasets. https://laion.ai/blog/laion-5b/. Accessed 15 October 2025.

Bender, E M., Gebru, T., McMillan-Major, A., Shmitchell, S. (2021). On the dangers of stochastic parrots: Can language models be too big? In: *Proceedings of the 2021 ACM Conference on Fairness, Accountability, and Transparency,* (pp. 610–623). New York, NY: Association for Computing Machinery.

Benjamin, W (1935). The Work of Art in the Age of Mechanical Reproduction In: Arendt, H. (ed.) *Illuminations*. New York, NY: Schocken Books.

Blackley, S. & Howell, J. (2015). A STEM Narrative: 15 Years in the Making.A STEM Narrative: 15 Years in the Making. *Australian Journal of Teacher Education,40*(7), 102–112. https://ro.ecu.edu.au/cgi/viewcontent.cgi?article=2720&context=ajte.Accessed 15 October 2025.

Campbell, C., & Olteanu, A. (2024). The Challenge of Postdigital Literacy: Extending Multimodality and Social Semiotics for a New Age. *Postdigital Science and Education 6*, 572–594. https://doi-org.ezproxy.jyu.fi/10.1007/s42438-023-00414-8

Cave, S., Craig, C., Dihal, K., Dillon, S., Montgomery, J., Singler, B., Taylor, L. (2018). *Portrayals and perceptions of AI and why they matter.* London, UK: The Royal Society. https://doi.org/10.17863/CAM.34502

Chang, Y, Wang, Y. Y., Ku, Y. T. (2023). Influence of online STEAM hands-on learning on AI learning, creativity, and creative emotions. *Interactive Learning Environments, 32(*8), 4719–4738. https://doi.org/10.1080/10494820.2023.2205898

Colucci-Gray, L., Burnard, P., Gray, D., & Cooke, C. (2019). A critical review of STEAM (science, technology, engineering, arts, and mathematics). *Oxford research encyclopedia of education*. https://doi.org/10.1093/acrefore/9780190264093.013.398

Couldry, N., & Mejias, U. A. (2019). Data colonialism: Rethinking big data's relation to the contemporary subject. *Television & New media, 20*(4), 336–349. https://doi.org/10.1177/15274764187966

Crawford, K. (2021). *The Atlas of AI: Power, politics, and the planetary costs of Artificial Intelligence.* New Haven, CO: Yale University Press.

Creative Applications (2022). Unlearning Language – Using AI to be less machine-like. https://www.creativeapplications.net/project/unlearning-language-being-less-machine-like/

Daneshpour, H., & Kwegyir-Afful, E. (2022). Analysing transdisciplinary education: a scoping review. *Science & Education, 31*(4), 1047–1074. https://doi.org/10.1007/s11191-021-00277-0

de Freitas, E., Lupinacci, J., Pais, A. (2017). Science and technology studies× educational studies: Critical and creative perspectives on the future of STEM education. *Educational Studies 53*(6), 551–559. https://doi.org/10.1080/00131946.2017.1384730

D'Ignazio, C., Klein, L. (2020). *Data feminism*. Cambridge, MA: MIT Press.

Dissanayake, E. (2003) The core of art—Making special. *Journal of the Canadian Association for Curriculum Studies 1*(2). https://jcacs.journals.yorku.ca/index.php/jcacs/article/download/16856/15662. Accessed 15 October 2025.

Durall, E., Carter, C., Burns, K. (2022). Transdisciplinary education and innovation through STEAM. In: Proceedings of the Mini-Conference on *Transdisciplinary Research and Design* (TRaD 2022): 14th February 2022, University of Oulu (online). Oulun yliopisto, Oulu. https://oulurepo.oulu.fi/handle/10024/34183. Accessed 15 Octiber 2025.

Ecker, U. K., Lewandowsky, S., Chang, E. P., Pillai, R. (2014). The effects of subtle misinformation in news headlines. *Journal of Experimental Psychology: Applied 20*(4), 323–335. https://doi.org/ 10.1037/xap0000028

European Parliament (2023). *Artificial intelligence, democracy and elections.* https://www.journalofdemocracy.org/articles/the-freedom-house-survey-for-2020-democracy-in-a-year-of-crisis/. Accessed 15 October 2025.

Fagerlund, J., Leino, K., Kiuru, N., Niilo-Rämä, M. (2022). Finnish teachers' and students' programming motivation and their role in teaching and learning computational thinking. *Frontiers in Education 7*, 948783. https://doi.org/10.3389/feduc.2022.948783

Fawns, T. (2023). Postdigital education. In Jandrić, P. (Ed) *Encyclopedia of postdigital science and education* (pp. 1–11). Cham: Springer Nature Switzerland. https://doi.org/10.1007/978-3-031-35469-4_52-1

Fawns, T., Ross, J., Carbonel, H., Noteboom, J., Finnegan-Dehn, S., & Raver, M. (2023). Mapping and tracing the postdigital: Approaches and parameters of postdigital

research. *Postdigital Science and Education, 5*(3), 623–642. https://doi.org/10.1007/s42438-023-00391-y

Frankfurt, H. (2005). *On bullshit.* Princeton, NJ: Princeton University Press.

Green, B., & Beavis, C. (Eds.) (2012). *Literacy in 3D: An integrated perspective in theory and practice*. Victoria, AU: ACER Press.

Heikkilä, M. (2023). This new data poisoning tool lets artists fight back against generative AI. *MIT Technology Review*. https://www.technologyreview.com/2023/10/23/1082189/data-poisoning-artists-fight-generative-ai/. Accessed 15 October 2025.

Heybach, J., Pickup, A. (2017). Whose STEM? Disrupting the gender crisis within STEM *Educational Studies 53*(6), 614-627. https://doi.org/10.1080/00131946.2017.1369085

Hood, E. J., Lewis, T. E. (2021). 'Oohing and ahhing': The power of thin(g)king in art education research. *International Journal of Education Through Art 17*(2), 223–233. https://doi.org/10.1386/eta_00062_1

How, M. L., & Hung, W. L. D. (2019). Educating AI-thinking in science, technology, engineering, arts, and mathematics (STEAM) education. *Education Sciences, 9*(3), 184. https://doi.org/10.3390/educsci9030184

Huang, J., Han, A., Villanueva, A., Liu, Z., Zhu, Z., Ramani, K., & Peppler, K (2024). Deepening children's STEM learning through making and creative writing. *International Journal of Child-Computer Interaction, 40*, 100651. https://doi.org/10.1016/j.ijcci.2024.100651

Huang, W., Hew, K. F., Fryer, L. K. (2021) Chatbots for language learning—Are they really useful? A systematic review of chatbot-supported language learning. *Journal of Computer Assisted Learning 38*(1), 237–257. https://doi.org/10.1111/jcal.12610

Huang, X., & Qiao, C. (2024). Enhancing computational thinking skills through artificial intelligence education at a STEAM high school. *Science & Education, 33*(2), 383–403. https://doi.org/10.1007/s11191-022-00392-6

Hubbert, J. (2024). 70+ women in technology. https://explodingtopics.com/blog/women-in-tech#sources. Accessed 17 October 2025.

Hsu, T. C., Abelson, H., Lao, N., & Chen, S. C. (2021). Is it possible for young students to learn the A. I-STEAM application with experiential learning? *Sustainability, 13*(19), 11114. https://doi.org/10.3390/su131911114

Jandrić, P. (2025). Postdigital transdisciplinarity: Towards a new language of possibility. *Communiar: Revista de Imagen, Artes y Educación Crítica y Social, 13*, 20–33.

Jandrić, P., Knox, J., Besley, T., Ryberg, T., Suoranta, J., & Hayes, S. (2018). Postdigital science and education. *Educational Philosophy and Theory, 50*(10), 893–899. https://doi.org/10.1080/00131857.2018.1454000

Jiang, J., Vetter, M. & Lucia, B. Toward a 'more-than-digital' AI literacy: Reimagining agency and authorship in the postdigital era with ChatGPT. *Postdigital Science and Educaction 6*, 922–939 (2024). https://doi.org/10.1007/s42438-024-00477-1

Jiang, Y., Li, X., Luo, H. (2022). Quo vadis artificial intelligence? *Discover Artificial Intelligence* 2(4). https://doi.org/10.1007/s44163-022-00022-8

Kalpokas, I. (2023). Work of art in the Age of Its AI Reproduction. *Philosophy & Social Criticism*. *51*(8), 1268–1286. https://doi.org/10.1177/01914537231184490

Kamegne, Y., Owusu, E., & Oladapo, H. (2025, May). Alexa, why can't you hear our accents: Cross cultural studies on the inclusivity of voice recognition systems. *In International conference on Human-Computer Interaction* (pp. 62–71). Cham: Springer Nature Switzerland.

Katzenbach, C., & Ulbricht, L. (2019). Algorithmic governance. *Internet Policy Review, 8*(4), 1–18. https://doi.org/10.14763/2019.4.1424

Kauffman, K., Williams, A (2023). Turk wars: How AI threatens the workers who fuel it. *Stanford Social Innovation Review*. https://ssir.org/articles/entry/ai-workers-mechanical-turk. Accessed 15 October 2025.

Knochel, A. D. (2023). Midjourney Killed the Photoshop Star: Assembling the Emerging Field of Synthography. *Studies in Art Education 64*(4), 467–481. https://doi.org/10.1080/00393541.2023.2255085

Kreinsen, M., Schulz, S. (2021). Students' Conceptions of Artificial Intelligence. In: WiPSCE '21*: Proceedings of the 16th Workshop in Primary and Secondary Computing Educatio*n, (pp. 1–2). New York , NY: Association for Computing Machinery. https://doi.org/10.1145/3481312.3481328

Lagioia, F., Rovatti, R., Sartor, G. (2023) Algorithmic fairness through group parities? The case of COMPAS-SAPMOC. *A.I & Society 38*(2), 459–478. https://doi.org/10.1007/s00146-022-01441-y

Lacković, N., Olteanu, A. & Campbell, C. Postdigital Literacies in Everyday Life and Pedagogic Practices. *Postdigital Science and Educaction 6*, 796–820 (2024). https://doi-org.ezproxy.jyu.fi/10.1007/s42438-024-00500-5

Land, M. H. (2013). Full STEAM ahead: The benefits of integrating the arts into STEM. *Procedia Computer Science, 20*, 547–552.

Larson, J., Mattu, S., Kirchner, L., Angwin, J. (2016). How we analyzed the COMPAS recidivism algorithm. *ProPublica.* https://www.propublica.org/article/how-we-analyzed-the-compas-recidivism-algorithm. Accessed 15 October 2025.

Lee Taylor, B. (2023). Long hours and low wages: the human labour powering AI's development. *The Conversation*. https://theconversation.com/long-hours-and-low-wages-the-human-labour-powering-ais-development-217038. Accessed 15 October 2025.

Lehtiniemi, T., Ruckenstein, M. (2021) Prisoners Training AI. Ghosts, humans and values in data labour. In: Pink, S., Berg, M., Lupton, D., Ruckenstein, M. (eds.) *Everyday Automation*, (pp. 185–199). London, UK: Routledge.

Luitse, D., & Denkena, W. (2021). The great transformer: Examining the role of large language models in the political economy of AI. *Big Data & Society, 8(*2), https://doi.org/10.1177/20539517211047734

Marton, F. (1981). Phenomenography—describing conceptions of the world around us. *Instructional Science 10*(2), 177–200. https://doi.org/10.1007/bf00132516

Matuk, C. (2023). Speculative design for stem learning. In Diamond, J. & Rosenfeld, S (Eds.). *Amplifying informal science learning* (pp. 293–299). London, UK: Routledge.

Mei, Q., Xie, Y., Yuan, W., Jackson, M. O. (2024). A Turing test of whether AI chatbots are behaviorally similar to humans. *Economic Sciences 121*(9), e2313925121. https://doi.org/10.1073/pnas.2313925121

Meltzoff, A (1995). Understanding the intentions of others: Re-enactment of intended acts by 18-month-old children. *Developmental Psychology 31*(5), 838–850. https://doi.org/10.1037/0012-1649.31.5.838

Mertala, P., Fagerlund, J. (2024) Finnish 5th and 6th graders' misconceptions about artificial intelligence. *International Journal of Child-Computer Interaction 39*, 100630. https://doi.org/10.1016/j.ijcci.2023.100630

Mertala, P., Fagerlund, J., Calderon, O. (2022). Finnish 5th and 6th grade students' pre-instructional conceptions of artificial intelligence (AI) and their implications for A. I literacy education. *Computers and Education: Artificial Intelligence 3*, 100095. https://doi.org/10.1016/j.caeai.2022.100095

Milara, I. S., Pitkänen, K., Laru, J., Iwata, M., Orduña, M. C., & Riekki, J. (2020). STEAM in Oulu: Scaffolding the development of a Community of Practice for local educators around STEAM and digital fabrication. *International Journal of Child-Computer Interaction, 26*, 100197. https://doi.org/10.1016/j.ijcci.2020.100197

Moore, M. (2021). Speech recognition for individuals with voice disorders. In *Multimedia for accessible human computer interfaces* (pp. 115–144). Cham: Springer International Publishing.

Moore, T. J., Stohlmann, M. S., Wang, H. H., Tank, K. M., Glancy, A. W., & Roehrig, G. H. (2014). Implementation and integration of engineering in K-12 STEM education. In *Engineering in pre-college settings: Synthesizing research, policy, and* practices (pp. 35-60). West Lafayette, IN: Purdue University Press.

Morari, M. (2023). Integration of the arts in STEAM learning projects. *Review of Artistic Education, 26,* 262–277.

Morozov, E. (2013). *To save everything, click here: The folly of technological solutionism*. New York, NY: PublicAffairs.

Noe, A. (2015). *Strange tools: Art and human nature.* New York, NY: Hill Wang.

O'Neil, C. (2016). *Weapons of math destruction: How big data increases inequality and threatens democracy*. London, UK: Penguin books.

Pangrazio, L. (2026). The (im)possibility of AI literacy. *Learning, Media and Technology, 51*(1), 1–7. https://doi.org/10.1080/17439884.2026.2615553

Pangrazio, L., Sefton-Green, J. (2020). The social utility of 'data literacy'. *Learning, Media and Technology 45*(2), 208–220. https://doi.org/10.1080/17439884.2020.1707223

Pennington, D., Ebert-Uphoff, I., Freed, N., Martin, J., & Pierce, S. A. (2020). Bridging sustainability science, earth science, and data science through interdisciplinary education. *Sustainability Science, 15*(2), 647–661. https://doi.org/10.1007/s11625-019-00735-3

Perignat, E., Katz-Buonincontro, J. (2019). STEAM in practice and research: An integrative literature review. *Thinking Skills and Creativity 31*, 31–43. https://doi.org/10.1016/j.tsc.2018.10.002

Perrigo, B. (2024). Exclusive: OpenAI used Kenyan workers on less than $2 per hour to make ChatGPT less toxic. *Time*. https://time.com/6247678/openai-chatgpt-kenya-workers/. Accessed 15 October 2025.

Rapanta, C., Bhatt, I., Bozkurt, A., Chubb, L. A., Erb, C., Forsler, I., ... & Jandrić, P. (2025). Critical GenAI literacy: Postdigital configurations. *Postdigital Science and Education, 7,* 1296–1333. https://doi.org/10.1007/s42438-025-00573-w

Ravi, S. (2023). Large language models — LLM's simple explanation for kids. *Medium.* https://medium.com/@sandhiyawor/large-language-models-llms-simple-explanation-for-kids-e7a92120264f. Accessed 15 October 2025.

Razi, A., & Zhou, G. (2022). STEM, iSTEM, and STEAM: What is next? *International Journal of Technology in Education, 5(*1), 1–29. https://scholar.uwindsor.ca/educationpub/73. Accessed 15 October 2025.

Relmasira, S. C., Lai, Y. C., Donaldson, J. P. (2023). Fostering AI literacy in elementary Science, Technology, Engineering, Art, and Mathematics (STEAM) education in the age of generative AI. *Sustainability 15*(18), 13595. https://doi.org/10.3390/su151813595

Repucci, S., Slipowitz, A. (2021). Democracy in a year of crisis. *Journal of Democracy 32*. https://www.journalofdemocracy.org/articles/the-freedom-house-survey-for-2020-democracy-in-a-year-of-crisis/. Accessed 15 October 2025.

Sainsbury, L (2007). *The Race to the top: A Review of government's science and innovation policies*. https://www.rsc.org/images/sainsbury_review051007_tcm18-103118.pdf. Accessed 15 October 2025.

Sanders, M. (2009). Stem, stem education, stemmania. *The Techmology Teacher*, December/January, 2009, 20–26. https://vtechworks.lib.vt.edu/server/api/core/bitstreams/b5f37b87-c914-4e5a-8abc-f9b491dc2e36/content. Accessed 15 October 2025.

Sarker, I. (2021). Machine learning: Algorithms, real-world applications and research directions. *S. N Computer Science* 2(3), 160. https://doi.org/10.1007/s42979-021-00592-x

Shan, S., Ding, W., Passananti, J., Wu, S., Zheng, H., Zhao, B. Y. (2023). Prompt-specific poisoning attacks on text-to-image generative models. *arXiv*. http://arxiv.org/abs/2310.13828v2. Accessed 15 October 2025.

Sharp, S. (2022). He's bigger than Picasso on AI platforms, and he hates it. https://hyperallergic.com/766241/hes-bigger-than-picasso-on-ai-platforms-and-he-hates-it/. Accessed 15 October 2025.

Schneider, B. (2022). Multilingualism and AI: The regimentation of language in the age of digital capitalism. *Signs and Society, 10*(3), 362–387.

Skjuve, M., Følstad, A., Fostervold, K. I., Brandtzaeg, P. B. (2021)- My chatbot companion-a study of human-chatbot relationships. *International Journal of Human-Computer Studies 149,* 102601. https://doi.org/10.1016/j.ijhcs.2021.102601

Slotte Dufva, T., Mertala, P. (2021). Sähköä ja alkemiaa: tekoälydiskurssit Yleisradion verkkoartikkeleissa. *Media ja Viestintä 44*(1), 95–115. https://doi.org/10.23983/mv.107302

Starkey, A. (2022). Bruce Springsteen, the Vietnam war and 'Born in the U.S.A.'. https://faroutmagazine.co.uk/vietnam-war-bruce-springsteen-born-in-the-usa/. Accessed 15 October 2025.

Stokel-Walker, C. (2022). This couple is launching an organization to protect artists in the AI era. https://www.inverse.com/input/culture/mat-dryhurst-holly-herndon-artists-ai-spawning-source-dall-e-midjourney. Accessed 15 October 2025.

Tomašev, N., Cornebise, J., Hutter, F., Mohamed, S., Picciariello, A., Connelly, B., Clopath, C. (2020). AI for social good: unlocking the opportunity for positive impact. *Nature Communications 11*(1), 2468. https://doi-org.ezproxy.jyu.fi/10.1038/s41467-020-15871-z

Vartiainen, H., Kahila, J., Tedre, M., López-Pernas, S., Pope, N. (2024). Enhancing children's understanding of algorithmic biases in and with text-to-image generative AI. *New Media & Society, 27*(9), 5342–5368.. https://doi.org/10.1177/14614448241252820

Velander, J., Otero, N., & Milrad, M. (2024). What is critical (about) AI literacy? Exploring conceptualizations present in AI literacy discourse. In Buch, A., Lindberg, Y., Cerratto Pargman, T. (eds) *Framing futures in postdigital education: Critical concepts for data-driven practices* (pp. 139–160). Cham: Springer Nature.

Vora, K., McCullough, S., & Giordano, S. (2022). Asking different questions in STEM research: Feminist STS approaches to STEM pedagogy. *ADVANCE Journal, 3*(1). https://www.advancejournal.org/article/33675-asking-different-questions-in-stem-research-feminist-sts-approaches-to-stem-pedagogy. Accessed 15 October 2025.

Vygotsky, L. S. (1987). The Collected works of L. S. Vygotsky: The fundamentals of defectology, vol. 2. Cham: Springer.

Wallach, W., Allen, C. (2009). Can (ro)bots be moral? In: Wallach, W., Allen, C (eds.*) Moral machines: Teaching robots right from wrong,* (pp. 55–71). London, UK: Oxford University Press. https://doi.org/10.1093/acprof:oso/9780195374049.003.0005

Walter, Y. (2024). Embracing the future of Artificial Intelligence in the classroom: the relevance of AI literacy, prompt engineering, and critical thinking in modern education. *International Journal of Educational Technology in Higher Education, 21*(1), 15. https://doi.org/10.1186/s41239-024-00448-3

Weber-Guskar, E. (2021) How to feel about emotionalized artificial intelligence? When robot pets, holograms, and chatbots become affective partners. *Ethics and Information Technology 23*, 601–610. https://doi.org/10.1007/s10676-021-09598-8

White, D. W. (2014). What is STEM education and why is it important. *Florida Association of Teacher Educators Journal, 1*(14), 1–9. http://www.fate1.org/journals/2014/white.pdf. Accessed 15 October 2025.

White, J., Fu, Q., Hays, S., Sandborn, M., Olea, C., Gilbert, H., Elnashar, A., Spencer-Smith, J., Schmidt, D C. (2023). A prompt pattern catalog to enhance prompt engineering with ChatGPT. *arXiv*. https://arxiv.org/abs/2302.11382. Accessed 15 October 2025.

Wilde, O. (1891). *Intentions*. Malton, UK: Methuen and Co.

Yoon, J., & Olsen, A. (2023). Promoting science affinities through a video project in a science, technology, and society (STS) learning approach. *International Journal of Education in Mathematics, Science, and Technology (IJEMST), 11*(4), 1073–1093. https://doi.org/10.46328/ijemst.3049

Zouda, M., El Halwany, S., Bencze, L. (2023). Science and technology studies informing STEM education: Possibilities and dilemmas. In: *Challenges in science education: Global perspectives for the future*, (pp. 201-227). Cham: Springer.